\documentclass[conference]{IEEEtran}
\IEEEoverridecommandlockouts
\usepackage{cite}
\usepackage{amsmath,amssymb,amsfonts}
\usepackage{algorithmic}
\usepackage{graphicx}
\usepackage{textcomp}
\usepackage{multicol}
\usepackage{multirow}
\usepackage{afterpage}
\usepackage{siunitx}
\usepackage{nccmath}
\usepackage{array}
\usepackage{makecell}
\usepackage{hhline}
\usepackage{booktabs}
\usepackage{xcolor}
\newcommand{\argmin}{\mathop{\mathrm{arg\,min}}}
\usepackage{newtxmath}
\def\BibTeX{{\rm B\kern-.05em{\sc i\kern-.025em b}\kern-.08em
    T\kern-.1667em\lower.7ex\hbox{E}\kern-.125emX}}

\usepackage{stackengine}
\def\delequal{\mathrel{\ensurestackMath{\stackon[1pt]{=}{\scriptstyle\Delta}}}}

\begin{document}
\title{Evaluation of {RF} Fingerprinting-Aided {RSS}-Based Target Localization for Emergency Response}

\makeatletter
\newcommand{\linebreakand}{%
  \end{@IEEEauthorhalign}
  \hfill\mbox{}\par
  \mbox{}\hfill\begin{@IEEEauthorhalign}
}
\makeatother

\author{\IEEEauthorblockN{Halim Lee}
\IEEEauthorblockA{\textit{School of Integrated Technology} \\
\textit{Yonsei University}\\
Incheon, Korea \\
halim.lee@yonsei.ac.kr} 
\and
\IEEEauthorblockN{Taewon Kang}
\IEEEauthorblockA{\textit{School of Integrated Technology} \\
\textit{Yonsei University}\\
Incheon, Korea \\
taewon.kang@yonsei.ac.kr}
\and
\IEEEauthorblockN{Suhui Jeong}
\IEEEauthorblockA{\textit{School of Integrated Technology} \\
\textit{Yonsei University}\\
Incheon, Korea \\
ssuhui@yonsei.ac.kr}
\linebreakand
\IEEEauthorblockN{Jiwon Seo} 
\IEEEauthorblockA{\textit{School of Integrated Technology} \\
\textit{Yonsei University}\\
Incheon, Korea \\
jiwon.seo@yonsei.ac.kr}
}

\maketitle

\begin{abstract}
Target localization is essential for emergency dispatching situations. Maximum likelihood estimation (MLE) methods are widely used to estimate the target position based on the received signal strength measurements. However, the performance of MLE solvers is significantly affected by the initialization (i.e., initial guess of the solution or solution search space). To address this, a previous study proposed the semidefinite programming (SDP)-based MLE initialization. However, the performance of the SDP-based initialization technique is largely affected by the shadowing variance and geometric diversity between the target and receivers. In this study, a radio frequency (RF) fingerprinting-based MLE initialization is proposed. Further, a maximum likelihood problem for target localization combining RF fingerprinting is formulated. In the three test environments of open space, urban, and indoor, the proposed RF fingerprinting-aided target localization method showed a performance improvement of up to 63.31\% and an average of 39.13\%, compared to the MLE algorithm initialized with SDP. Furthermore, unlike the SDP-MLE method, the proposed method was not significantly affected by the poor geometry between the target and receivers in our experiments.
\end{abstract}


\section{Introduction}
Timely and precise positioning is critical for emergency response. In 2018, the Federal Communications Commission (FCC) estimated that ``10,000 lives could be saved each year if the emergency dispatching system (9-1-1) could get help one minute sooner to those calling for emergency assistance'' \cite{NENA}. For public safety, several countries, including the United States and Europe, are enacting standards and requirements to ensure the emergency positioning quality of mobile phones \cite{FCC15, EU14}. For instance, the FCC is currently forcing commercial mobile radio service (CMRS) providers to satisfy the positioning requirements of a vertical accuracy of 3 m and a horizontal accuracy of 50 m for 80\% of all indoor 911 calls \cite{FCC15}.

In an open-sky environment, meter-level positioning accuracy can be achieved by mobile phones using global navigation satellite systems (GNSS) \cite{Lee17:Monitoring, Dabove19:Towards, Li19:Characteristics, Yoon20:An, Bang13:Methodology}. However, the accuracy of GNSS can be drastically degraded in urban or indoor areas \cite{DeAngelis12:GNSS, Soloviev10:Tight, Lee20202347, Opshaug01:GPS, Dong18:ViNav, Agarwal02:Algorithms, Lee22:Nonlinear}.
Furthermore, GNSS is vulnerable to radio frequency (RF) interference \cite{Rhee21:Enhanced, Kim22:First, Lee22:SFOL, Park2021919, Son20191828} and ionospheric anomalies \cite{Lee22:Optimal, Sun21:Markov, Yoon14:Medium}. 
Therefore, an alternative E911 positioning method is required in GNSS-denied environments.

Target localization methods can be used to estimate the position of the emergency caller in GNSS-denied environments. Target localization (i.e., sensor localization or passive emitter localization) has been mainly studied in the wireless sensor networks (WSN) field \cite{Vaghefi12:Cooperative, Ouyang10:Received, Li07:Collaborative, Sheng04:Maximum}. The purpose of target localization in WSNs is to estimate the positions of sensor nodes through noisy measurements measured by anchor nodes with known positions. As opposed to anchor nodes assumed to be static in WSNs, target localization methods assuming movable receivers, such as unmanned aerial vehicles (UAVs), are also being actively studied \cite{Jeong2020958, Wang18:Performance, Uluskan20:Noncausal, Morales17:Optimal} for both civilian (e.g., search and rescue) and military (e.g., surveillance) applications.

Various measurements, such as received signal strength (RSS) \cite{Vaghefi12:Cooperative, Li07:Collaborative, Ouyang10:Received, Sheng04:Maximum, Jeong2020958, Wang18:Performance, Uluskan20:Noncausal}, time-of-arrival (TOA) \cite{Uluskan20:Noncausal, Morales17:Optimal, Yang09:Efficient, Chan06:Exact, Gaber14:A, Shamaei21:Receiver, Wang20:Performance, Wang20:Multipath, Rosado19:Physical}, direction-of-arrival (DOA) \cite{Wang13:Convex, Xu17:Optimal, Abdallah21:Multipath}, and Doppler shift \cite{Wei09:Multidimensional, Yang16:Moving}, can be used for target localization. Among these measurements, the RSS is attractive because it can be measured with relatively simple hardware and software resources. Therefore, this study focuses on RSS-based target localization.

Maximum likelihood estimation (MLE) is a promising method for estimating the position of a target using RSS measurements, and has been used in target localization studies \cite{Vaghefi12:Cooperative, Li07:Collaborative, Ouyang10:Received, Sheng04:Maximum, Jeong2020958, Wang18:Performance, Uluskan20:Noncausal}. However, because the cost function of MLE is nonlinear and nonconvex in RSS-based target localization problems, appropriate initialization has a large impact on MLE performance \cite{Vaghefi12:Cooperative, Li07:Collaborative, Ouyang10:Received, Sheng04:Maximum}. If the initial values of iterative methods, such as the gradient descent method \cite{Baldi95:Gradient, Garg12:An} and Newton's method \cite{Galantai00:The}, are not close enough to the global minimum, the algorithm may converge to the local minimum or saddle point and cause a large estimation error. Similarly, in a grid search-based MLE, the search area must be initialized when there is insufficient time to search all grids. In this case, the estimation performance varied depending on the initialization \cite{Heidenreich13:Fast}.

To resolve the initialization problem in MLE, convex relaxation and linear estimation have been used to specify the initial points in the literature \cite{Vaghefi12:Cooperative, Li07:Collaborative, Ouyang10:Received, Sheng04:Maximum}. Convex- or linear-estimation-based initialization can be useful if the number of receivers and geometric diversity between the target and receiver are sufficient \cite{Sadeghi20:Optimal, Bishop10:Optimality}. 
However, the number of receivers (e.g., emergency dispatchers or drones) and time required to secure geometric diversity might be insufficient during an emergency. Furthermore, owing to the large shadowing variance in complex environments, these methods have low position estimation accuracy in urban or indoor areas. Therefore, an alternative method that can effectively initialize the MLE is required for emergency situations.

In this study, we suggested the method of MLE initialization through RF fingerprinting. RF fingerprinting is a localization method that identifies the position based on the received signal characteristics (e.g., RSS and Cell ID), which are ``fingerprints'' of the signals that are expected to be received at a corresponding location \cite{Huang17:A, Hu18:Experimental, Tian16:Performance, Djosic21:Fingerprinting, Lee22:Automatic, Bai21:Distance}. 
The search space of MLE was initialized using RF fingerprinting, and the final position solutions were obtained based on our RF fingerprinting-combined maximum likelihood (ML) problem formulation. 

There are two advantages of adding RF fingerprinting modality to target localization compared to existing convex- or linear-estimation-based MLE initialization. 
First, RF fingerprinting-based initialization is less affected by large-scale fading (i.e., shadowing), contrary to the existing initialization methods that are affected by both large- and small-scale fading. 
The pre-recorded RF fingerprinting map stores averaged RSS measurements collected over a long period of time. Hence, the effect of large-scaling fading contained in the RF fingerprinting map is already mitigated by the average filter compared to the case of real-time RSS measurements. 
Second, the RF fingerprinting modality increases geometric diversity for localization. The proposed method uses both uplink (UL) and downlink (DL) signals, contrary to the existing methods that rely only on the UL signal. 
The DL signal link between the target and LTE base station improves the localization geometry compared to the case of using UL signal alone between the target and receiver. 

The RF fingerprinting-aided target localization system proposed in this study consists of the following three parts:

\begin{itemize}
    \item The \textbf{target}, where the position should be estimated, sends a rescue request to the localization server. It also measures the RSS of the DL signal from the LTE base station and sends it to the localization server.
    \item The \textbf{receivers} measure the RSS of the UL signal transmitted from the target and send it to the localization server.
    \item The \textbf{localization server} estimates the target position. The localization server performs RF fingerprinting based on the RSS measurements of the DL signal for MLE initialization and estimates the position of the target using the RSS measurements of both DL and UL signals.
\end{itemize}
To  consider realistic emergency situations, we evaluated the performance of RF fingerprinting-aided target localization through experiments in three environments: open space, urban, and indoor areas.

The remainder of this paper is organized as follows. The RSS-based target-localization problem is described in Section \ref{sec:RSS-BasedTargetLocalization}. Section \ref{sec:FP-AidedTargetLocalization} describes the RF fingerprinting-aided target localization method. Section \ref{sec:TestEnvironments} introduces the test environments, and Section \ref{sec:Evaluation} presents the evaluation results based on experiments conducted in the test environments. Finally, the conclusions are presented in Section \ref{sec:Conclusion}.

\section{RSS-Based Target Localization} \label{sec:RSS-BasedTargetLocalization}

This section introduces the RSS-based target localization problem using only the RSS measurements of the UL signal. Using the RSS measured by $N$ receivers, the target's 2D position $\mathbf{x}=[x_t,y_t]^\mathsf{T}$ is estimated. According to the log-distance path loss model, the RSS (in dBm) measured by the $i$-th receiver, $P_{\textrm{UL},i}$, can be modeled as \cite{Vaghefi12:Cooperative, Li07:Collaborative, Ouyang10:Received, ITUR}:

\begin{equation} \label{eq:Path loss model}
\begin{split}
P_{\textrm{UL},i} &= P_{\textrm{UL},0} - 10 \beta \log_{10} \frac{d_i}{d_0} + n_i,\\
d_i &= \|\mathbf{x}-\mathbf{r}_i\|, \\
n_i &\sim \mathcal{N}(0, \sigma_\mathrm{dB}^2),
\end{split}
\end{equation}
where $P_{\mathrm{UL},0}$ (in dBm) is the signal power at the reference distance $d_0$ from the target ($d_0$ is assumed to be 1 m in this paper); $\beta$ is the path loss exponent that varies depending on the environment; $n_i$ is the log-normal shadowing term which can be modeled as Gaussian random variable with standard deviation of $\sigma_\mathrm{dB}$; $\mathbf{r}_i=[x_{r_i},y_{r_i}]^\mathsf{T}$ is the 2D position of $i$-th receiver; and $\|\cdot\|$ denotes $L^2$ norm. 
In (\ref{eq:Path loss model}), there are three unknowns that need to be estimated: the target's 2D position, $\mathbf{x}=[x_t,y_t]^\mathsf{T}$, and the signal power at the reference distance, $P_{\mathrm{UL},0}$. The signal power at the reference distance, $P_{\mathrm{UL},0}$, depends on the transmit power of the target.

\subsection{Maximum Likelihood Estimation}

The unknowns, $\boldsymbol{\thetaup} = [\mathbf{x}^\mathsf{T}, P_{\textrm{UL},0}]^\mathsf{T} = [x_t, y_t,  P_{\textrm{UL},0}]^\mathsf{T}$, can be estimated with the maximum likelihood estimator, as follows \cite{Vaghefi12:Cooperative, Ouyang10:Received}:

\begin{equation} \label{eq:MLE}
    \widehat{\boldsymbol{\thetaup}} = \argmin_{\boldsymbol{\thetaup}} \sum_{i}^{N} \left( P_{\textrm{UL},i} - P_{\textrm{UL},0} + 10 \beta \log_{10} d_i \right) ^2.
\end{equation}
Because (\ref{eq:MLE}) does not have a closed-form solution, it needs to be solved using numerical methods \cite{Baldi95:Gradient, Garg12:An, Galantai00:The}. However, as previously mentioned, appropriate initial values must be provided to efficiently determine the global minimum of this ML problem.

Among the techniques used to determine the initial value of the MLE, semidefinite programming (SDP) derived in \cite{Vaghefi12:Cooperative} was considered in this study. The SDP algorithm presented in \cite{Vaghefi12:Cooperative} can be used alone or together with MLE (i.e., the SDP algorithm can be used to initialize the MLE) to estimate the position of the target. In terms of positioning accuracy, the SDP-initialized MLE showed a better performance than both the SDP alone and the MLE initialized with the least-squares method \cite{Vaghefi12:Cooperative}.

\subsection{Semidefinite Programming}

In \cite{Vaghefi12:Cooperative}, the nonconvex and nonlinear ML problem of (\ref{eq:MLE}) was relaxed into an SDP optimization problem. The final SDP problem can be formulated as \cite{Vaghefi12:Cooperative}:

\begin{equation} \label{eq:SDP}
    \begin{split}
        \underset{\mathbf{x}, z, \alpha, h_i}{\text{Minimize}} \quad & \sum_{i}^{N} \left( h_i \lambda_i - \alpha \right)^2\\
        \text{Subject to} \quad & h_i = \begin{bmatrix} \mathbf{r}_i\\-1 \end{bmatrix} ^ \mathsf{T} \begin{bmatrix} \mathbf{I}_\mathrm{2} & \mathbf{x} \\ \mathbf{x}^\mathsf{T} & z \end{bmatrix} \begin{bmatrix} \mathbf{r}_i \\ -1 \end{bmatrix}, \\
        & \begin{bmatrix} \mathbf{I}_\mathrm{2} & \mathbf{x} \\ \mathbf{x}^\mathsf{T} & z \end{bmatrix} \succeq \mathbf{0}_\mathrm{3},
    \end{split}
\end{equation}
where $h_i \delequal d_i^2$; $\lambda_i \delequal 10^{P_{\textrm{UL},i}/5\beta}$; $\alpha \delequal 10^{P_{\textrm{UL},0}/5\beta}$; and $z$ is an auxiliary variable defined as $z=\mathbf{x}^\mathsf{T}\mathbf{x}$.

The cost function of the ML estimator in (\ref{eq:MLE}) is converted into a convex cost function in (\ref{eq:SDP}) after two steps: approximation using first-order Taylor series expansion and convex relaxation. Detailed SDP conversion procedures for the ML problem can be found in \cite{Vaghefi12:Cooperative}.

Solving the SDP problem of (\ref{eq:SDP}) can provide good initial values of the ML problem in (\ref{eq:MLE}) when the geometry between the receivers and the target is sufficiently diverse and the shadowing variance, $\sigma_\mathrm{dB}^2$ in (\ref{eq:Path loss model}), is sufficiently small. However, as previously mentioned, a more precise MLE initialization method is needed especially in emergencies. 
In emergency response situations, resources (i.e., time and the number of receivers) are likely insufficient to secure a good target-receiver geometry. 
Further, in complex environments such as urban and indoor areas, significant shadow fading could occur due to the building obstacles \cite{ITUR, Algans02:Experimental}. 
Considering that the majority of emergency calls occur in these areas \cite{FCC15}, a better MLE initialization method for urban and indoor areas is needed.

\section{RF Fingerprinting-Aided Target Localization} \label{sec:FP-AidedTargetLocalization}

This section proposes the RF fingerprinting-aided target localization method, which localizes the target using RSS measurements of both the UL and DL signals. The RF fingerprinting-aided target localization algorithm in this study includes two steps: initializing the MLE algorithm with RF fingerprinting, and estimating the position of the target using the proposed equation that combines the RF fingerprinting measurements into the ML problem in (\ref{eq:MLE}).

\subsection{MLE Initialization with RF Fingerprinting}

RF fingerprinting algorithms are generally divided into two phases: offline data training phase and online localization phase \cite{Huang17:A}. In the offline phase, a database that records the signal characteristics transmitted from stationary devices (e.g., WiFi access point, Bluetooth beacon, and LTE base station), which is an RF fingerprinting map, is built into the localization server. In this study, we created an RF fingerprinting map based on the RSS of DL signals transmitted from nearby LTE base stations. 
In the online phase, the position is estimated using pattern-matching algorithms by comparing the RSS measured in real time with the RF fingerprinting map. 

In this study, the $k$-nearest neighbor ($k$NN) algorithm was used for pattern matching. $k$NN is a widely used classification algorithm due to its simplicity \cite{Huang17:A, Hu18:Experimental, Tian16:Performance}. The $k$NN procedure used in this study was as follows. 
First, the RSS of the DL signals is stored in a training database with labels. The labels can be the position or category (e.g., name of the nearby building) of the corresponding signal collection point. 
Second, during the online phase, the RSS of the DL signal is measured in real-time and sent to the localization server. 
The localization server calculates the distance, which is the mean-squared error (MSE) between every RSS stored in the training database and the RSS measured in real-time. 
Then, $k$ data points with the smallest MSEs were selected. 
Finally, the final label was determined by the largest number of labels among the labels of the $k$ data points. 
The final label was used for the MLE initialization. 

If the final label is a position label, MLE is applied within a specific range of areas surrounding the position in the label. If the final label is a category label, MLE is applied within a predefined area that corresponds to the category in the label (e.g., a predefined area around the corresponding building in the label).

%After determining the final label, MLE was performed on the search space that corresponds to the label (e.g., a certain area around the point or a pre-segmented area).

%The estimated position was used as the initial value for the MLE algortihm. 

\subsection{Maximum Likelihood Estimation Combining RF Fingerprinting}

The RF fingerprinting-combined ML problem for estimating the final position solution of the target can be formulated as follows:

\begin{equation} \label{eq:MLEwithFP}
    \begin{split}
        \widehat{\boldsymbol{\thetaup}} = \argmin_{\boldsymbol{\thetaup}} \sum_{i}^{N} & \left( P_{\mathrm{UL},i} - P_{\mathrm{UL},0} + 10 \beta \log_{10} d_i \right) ^2 \\
        & + w \cdot \left( P_{\mathrm{DL}} - P_{\mathrm{map},  \textbf{x}} \right)^2 ,
    \end{split}
\end{equation}
where $w$ is a weighting factor; $P_{\mathrm{DL}}$ is the RSS of the DL signal, which is measured by the target terminal; and $P_{\mathrm{map},\mathbf{x}}$ is the RSS fingerprinting data recorded at $\mathbf{x}$. 
If there are no data recorded in the RF fingerprinting map at point $\mathbf{x}$, the data of the nearest point are used. In contrast to (\ref{eq:MLE}), where only RSS measurements of UL signals are considered, RF fingerprinting errors are also considered in (\ref{eq:MLEwithFP}) to improve the positioning performance.

\section{Test Environments} \label{sec:TestEnvironments}

The RF fingerprinting-aided target localization algorithm was evaluated in three different environments: open space, urban, and indoor environments.

\subsection{Open Space}

An open-space experiment was conducted at the stadium of Hanyang University, Seoul, Korea, as shown in Fig. \ref{fig:open_space}. A fingerprinting map was created by measuring the RSS of the DL signal from a nearby base station. The LTE base station was placed on the roof of the R\&D building at Hanyang University, which is approximately 150 m away from the stadium. A fingerprinting map was built by measuring the RSS at intervals of 10 m and covering the entire field. Each RSS measurement stored in the training database was labelled with a 2D position. A Samsung Galaxy S8+ was used as the target and receiver device, and an application that could measure the RSS developed by Hanyang University was installed on each device.

The receivers were placed as indicated by the yellow icons in Fig. \ref{fig:open_space}. The target terminal was moved and placed at seven points (red icons in Fig. \ref{fig:open_space}). For each placement, the target terminal measured the RSS of the DL signal transmitted from the base station. 
The target terminal transmitted the LTE band 28 UL signal with a bandwidth of 10 MHz (permission was granted to transmit LTE band 28 DL and UL signals within the Hanyang University demonstration complex). 
The receivers measured the RSS of the UL signal ten times at each target placement. Therefore, 70 RSS measurements of the UL signal transmitted from the target were recorded.

\begin{figure} 
    \centering 
    \includegraphics[width=\linewidth]{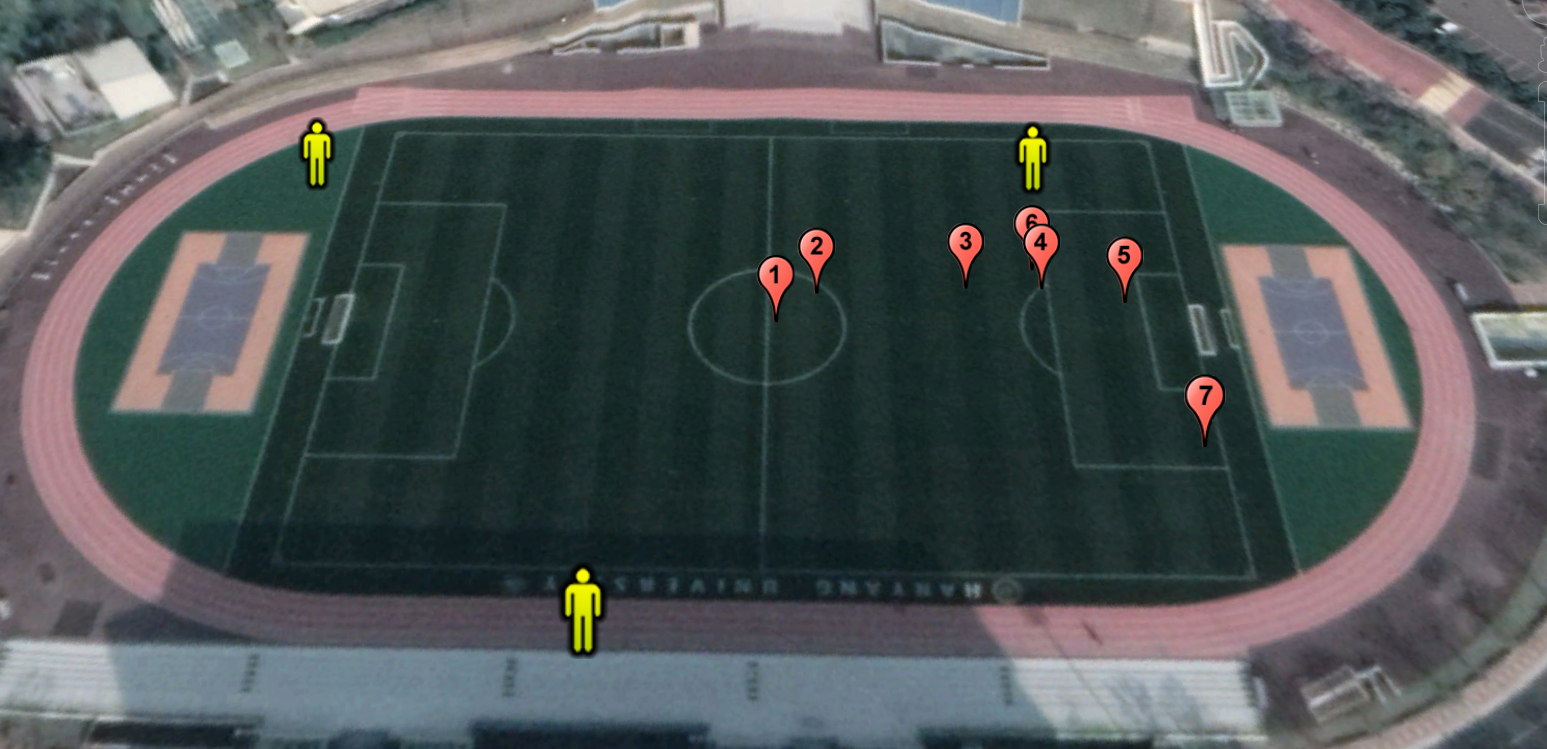}
    \caption{An open space test environment. The yellow icons represent the receivers and the red icons represent the positions of the target.} \label{fig:open_space}
\end{figure}

\subsection{Urban Environment}

Fig. \ref{fig:urban} shows the urban experimental environment at Yonsei University, Incheon, Korea. The fingerprinting data were collected from the spots marked with red dots. In the urban experiment, the RSS data were labelled with the nearby building names: General Education Hall, Libertas Hall A, and Veritas Hall C. The LTE base station was located as indicated by the blue mark in Fig. \ref{fig:urban}. To collect the LTE signal, PicoZed software-defined radio (SDR) with a WA700/2700 antenna from PulseLarsen was used.

\begin{figure} 
    \centering
    \includegraphics[width=\linewidth]{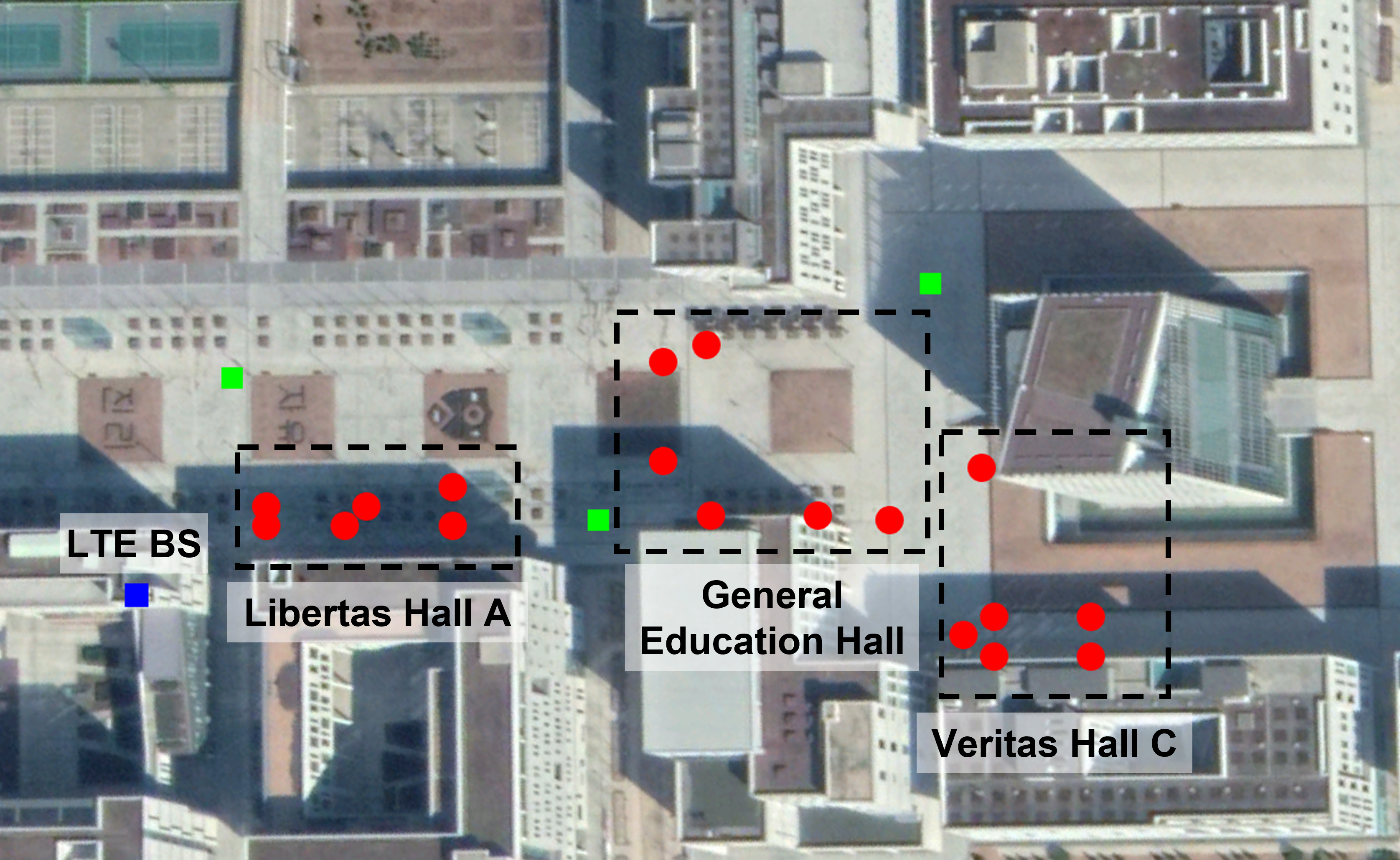}
    \caption{An urban test environment. The blue square represents the LTE base station. Fingerprinting data collection points are marked as red dots and the positions of the receivers are marked as green squares. The fingerprinting data were labeled by the name of nearby buildings: Libertas Hall A, General Education Hall, and Veritas Hall C.} \label{fig:urban}
\end{figure}

RSS measurements of the UL signal transmitted from the target were generated through simulations. During the simulation, the true position of the target was randomly selected from the 18 red dots in Fig. \ref{fig:urban}. The receivers were assumed to be fixed as the green squares in Fig. \ref{fig:urban}. There were 10 random target placements, and 100 simulated RSSs were generated for each target placement. Hence, 1,000 target localizations were performed. Referring to an urban-macro cell line-of-sight (LoS) scenario of the international telecommunication union-radiocommunication sector (ITU-R) channel model, the log-normal shadowing term, $\sigma_{\mathrm{dB}}$ in (\ref{eq:Path loss model}), was set to 4 dB \cite{ITUR}.

\subsection{Indoor Environment}

Fig. \ref{fig:indoor} shows the indoor experimental environment of the Veritas Hall C at Yonsei University. The fingerprinting data were collected on spots marked with red dots and stored with 2D position labels. The LTE signals transmitted from three repeaters located indoors were collected by the same equipment that was used in the urban experiments. 
During the experiments, the doors of all rooms were left open. The experiment was conducted for three days; the data from the first two days were used for the offline data training phase, and the data from the last day were used for the online localization phase.

As in the urban experiment, the RSS values of the UL signals were generated by simulation. The receivers were assumed to be fixed as the green squares in Fig. \ref{fig:indoor}. A total of 100 RSS data points were generated for all the points marked with red dots. Thus, 1,300 target localizations were performed. The log-normal shadowing term, $\sigma_{\mathrm{dB}}$ in (\ref{eq:Path loss model}), was set to 3 dB by referring to an indoor hotspot LoS scenario of ITU-R channel model \cite{ITUR}.

\begin{figure} 
    \centering
    \includegraphics[width=\linewidth]{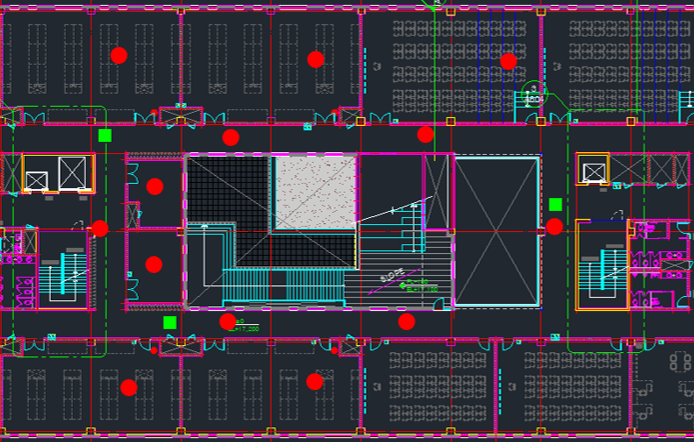}
    \caption{An indoor test environment. Fingerprinting data collection points are marked as red dots and the positions of the receivers are marked as green squares.} \label{fig:indoor}
\end{figure} 

\section{Test Results} \label{sec:Evaluation}

\subsection{Performance Improvement over the Existing Methods}

Fig. \ref{fig:Results} shows the 2D position error and cumulative distribution function (CDF) of the test results for open space, urban, and indoor environments. We compared our algorithm (i.e., fingerprinting-aided MLE, or FP-MLE) with randomly-initialized MLE (RAND-MLE) and SDP-initialized MLE (SDP-MLE). In the $k$NN algorithm, $k$ was set to 1 for open spaces and indoor environments and set to 5 for urban environments. In all the environments, $w$ was set to 0.01. 

A grid search-based MLE was implemented for every case. 
The range for the grid search was set to 15 m for open spaces and 10 m for indoor environments from the initial point in all algorithms. 
In the case of urban experiments, the range for the grid search was set to 30 m in RAND-MLE and SDP-MLE algorithms. For FP-MLE, the black-dashed area in Fig. \ref{fig:urban} that corresponded to the obtained label (i.e., building name) was used as the range for the grid search.

\begin{figure} 
    \centering
    \includegraphics[width=\linewidth]{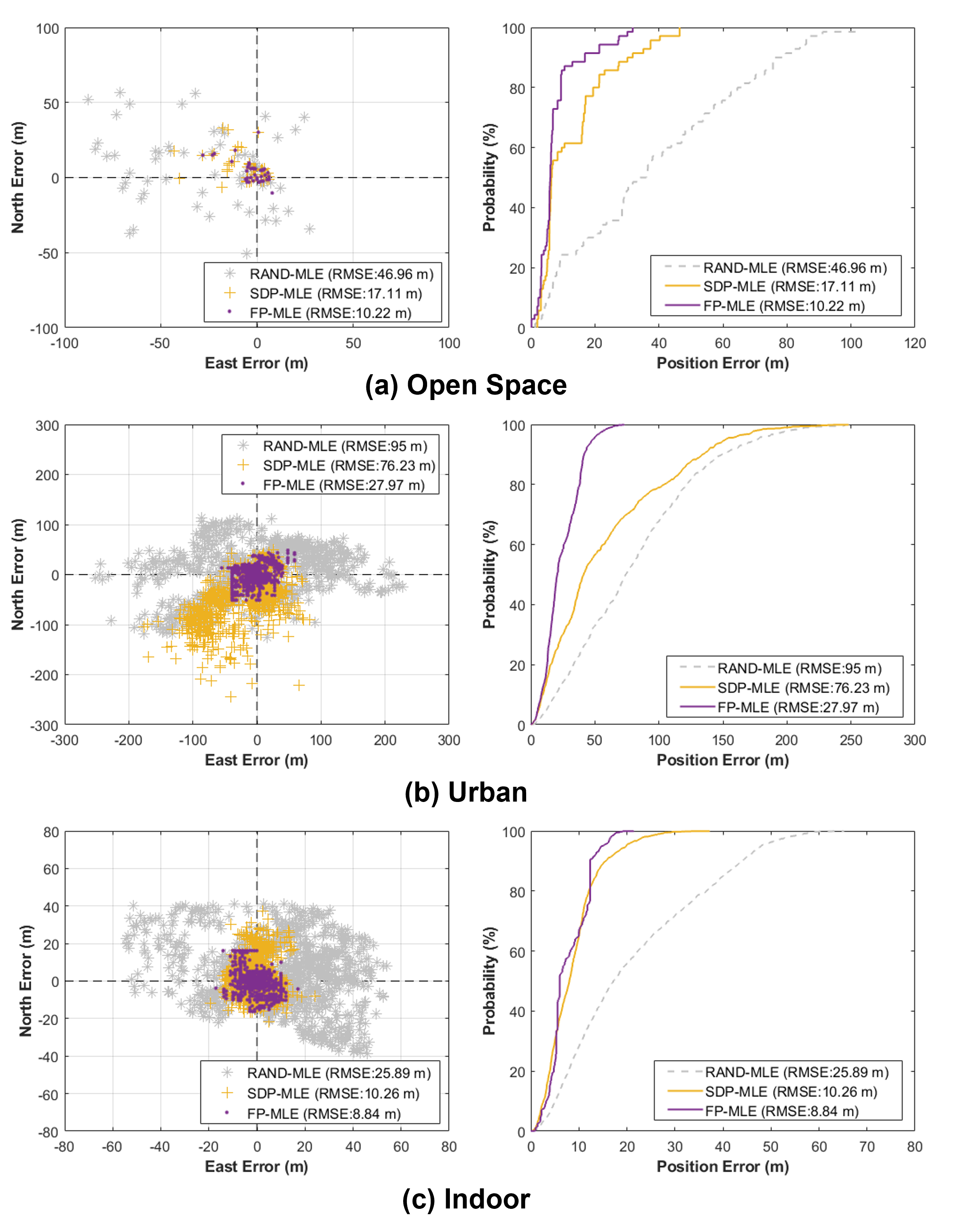}
    \caption{Test results for (a) open space, (b) urban, and (c) indoor environments.} \label{fig:Results}
\end{figure} 

Table \ref{tab:Performance} compares the root-mean-squared error (RMSE) and the 80th percentiles of the 2D positioning errors. Our algorithm showed better positioning performance in all cases compared with RAND-MLE and SDP-MLE. Compared with SDP-MLE, the RMSE of FP-MLE was smaller by 40.3\%, 63.3\%, and 13.8\% in the open space, urban, and indoor environments, respectively. In addition, the 80th percentiles of the 2D positioning error of FP-MLE were smaller than 50 m in all cases, which satisfied the FCC E911 requirement for horizontal accuracy. This suggests that FP-MLE can potentially be utilized for E911 positioning in contrast to RAND-MLE or SDP-MLE, which did not satisfy the corresponding requirement during our tests.

\begin{table} 
\centering
\caption{Performance comparison among three algorithms (unit: m)}
\label{tab:Performance}
\vspace{-4mm}
\begin{center}
{\renewcommand{\arraystretch}{1.4}
 \begin{tabular}[c]{>{\centering\arraybackslash}m{0.7cm}>{\centering\arraybackslash}m{0.8cm}>{\centering\arraybackslash}m{1.5cm} >{\centering\arraybackslash}m{1.5cm}>{\centering\arraybackslash}m{2.0cm}}
 \toprule
 \noalign{\vspace{-2pt}}
    {} & {} & \vspace{5pt}\thead{RAND-MLE} 
       & \vspace{5pt}\thead{SDP-MLE} 
       & \thead{FP-MLE\\(proposed)} \\
    \noalign{\vspace{0pt}}
\hline
\noalign{\vspace{2pt}}
\multirow{2}{*}{\thead{Open\\Space}} 
 & RMSE & 46.96 & 17.11 & 10.22 \\
 & 80\% & 64.50 & 19.44 & 9.43 \\
 \noalign{\vspace{1pt}}
 \midrule
 \multirow{2}{*}{Urban} 
 & RMSE & 95.00 & 76.23 & 27.97 \\
 & 80\% & 120.40 & 106.00 & 38.34 \\
 \noalign{\vspace{1pt}}
 \midrule
 \multirow{2}{*}{Indoor} 
 & RMSE & 25.89 & 10.26 & 8.84 \\
 & 80\% & 36.55 & 12.07 & 12.35 \\
 \noalign{\vspace{1pt}}
 \bottomrule
\end{tabular}}
\end{center}
\end{table}

\subsection{Performance in Poor Geometry Scenarios}

Scenarios with poor geometries between the target and receivers were tested for urban and indoor environments. In the poor geometry scenarios, it was assumed that three receivers were clustered within a 10 m radius on the upper left corner of the test environments. 
Table \ref{tab:Bad} compares the RMSE and 80th percentiles of the 2D positioning errors in poor geometry scenarios. Despite the poor geometric diversity between the target and receivers, FP-MLE still satisfied the 50 m horizontal positioning requirement in both urban and indoor environments. On the other hand, SDP-MLE showed almost the same positioning performance as RAND-MLE especially in an urban environment, where shadowing variance, $\sigma_{\mathrm{dB}}^2$ in (\ref{eq:Path loss model}), is large. This means that MLE initialization with the SDP algorithm is vulnerable to poor geometric diversity and large shadowing variance.

\begin{table} 
\centering
\caption{Performance comparison among three algorithms in poor geometry scenarios (unit: m)}
\label{tab:Bad}
\vspace{-4mm}
\begin{center}
{\renewcommand{\arraystretch}{1.4}
 \begin{tabular}[c]{>{\centering\arraybackslash}m{0.7cm}>{\centering\arraybackslash}m{0.8cm}>{\centering\arraybackslash}m{1.5cm} >{\centering\arraybackslash}m{1.5cm}>{\centering\arraybackslash}m{2.0cm}}
 \toprule
 \noalign{\vspace{-2pt}}
    {} & {} & \vspace{5pt}\thead{RAND-MLE} 
       & \vspace{5pt}\thead{SDP-MLE} 
       & \thead{FP-MLE\\(proposed)} \\
    \noalign{\vspace{0pt}}
\hline
\noalign{\vspace{2pt}}
 \multirow{2}{*}{Urban} 
 & RMSE & 101.44 & 100.26 & 35.29 \\
 & 80\% & 138.02 & 144.86 & 44.19 \\
 \noalign{\vspace{1pt}}
 \midrule
 \multirow{2}{*}{Indoor} 
 & RMSE & 22.18 & 15.43 & 12.70 \\
 & 80\% & 28.78 & 18.84 & 16.43 \\
 \noalign{\vspace{1pt}}
 \bottomrule
\end{tabular}}
\end{center}
\end{table}

\section{Conclusion} \label{sec:Conclusion}

This study investigated how performance can be improved when the RF fingerprinting modality is added to the RSS-based target localization method. Fingerprinting was utilized for MLE initialization, and a fingerprinting-combined ML problem was proposed. The 2D positioning performance of the fingerprinting-aided target localization method (i.e., FP-MLE) was compared with that of the MLE algorithms initialized using SDP (i.e., SDP-MLE) or randomly (i.e., RAND-MLE). 
The positioning performance of FP-MLE was superior to that of SDP-MLE and RAND-MLE in open space, urban, and indoor environments. In addition, FP-MLE satisfied the FCC's E911 horizontal accuracy requirement in all environments, unlike other algorithms. As a future study, more realistic channel models (e.g., channel models with additional path loss owing to walls) need to be considered and movable target scenarios can also be considered.

\section*{Acknowledgment}

The authors would like to thank the Communication System Laboratory of Hanyang University, Korea, for the outdoor data collection. 
This research was supported in part by the Institute of Information \& Communications Technology Planning \& Evaluation (IITP) grant funded by the Korea government (KNPA) (2019-0-01291) and in part by the Basic Science Research Program through the National Research Foundation of Korea (NRF) funded by the Ministry of Education (2021R1A6A3A13046688). 

\bibliographystyle{IEEEtran}
\bibliography{mybibfile, IUS_publications}

\end{document}